\documentclass[conference,a4paper]{IEEEtran}

\usepackage[utf8]{inputenc} 
\usepackage[T1]{fontenc}
\usepackage{url}
\usepackage{ifthen}
\usepackage{cite}
\usepackage[cmex10]{amsmath} 
\usepackage{amssymb} 
\usepackage{tikz}   
\usepackage{mathtools}  
\DeclarePairedDelimiter\floor{\lfloor}{\rfloor}                       

\newtheorem{theorem}{Theorem}
\newtheorem{proposition}{Proposition}

\begin{document}
\title{The Asymptotic Capacity of Private Search} 


\author{%
  \IEEEauthorblockN{Zhen Chen, Zhiying Wang and Syed Jafar}
  \IEEEauthorblockA{Center for Pervasive Communications and Computing (CPCC)\\
                    University of California Irvine\\
                    Irvine, CA 92697\\
                    Email: \{zhenc4, zhiying, syed\}@uci.edu}
}


\maketitle

\begin{abstract}
The private search problem is introduced, where a dataset comprised of $L$ i.i.d. records is replicated across $N$ non-colluding servers, each record takes values uniformly from an alphabet of size $K$, and a user  wishes to  search for all records that match a privately chosen value, without revealing any information about the chosen value to any individual server. The capacity of private search is the maximum number of bits of desired information that can be retrieved per bit of download. The asymptotic (large $K$) capacity of private search is shown to be $1-1/N$, even as the scope of private search is further generalized to allow approximate (OR) search over a number of realizations that  grows with $K$. The results are based on  the asymptotic behavior of a new converse bound for private information retrieval with arbitrarily dependent messages.
\end{abstract}


\section{Introduction}
\emph{Search} is among the most frequent operations performed on large datasets that are stored online. With privacy concerns increasingly taking center stage in online interactions, a private search functionality is highly desirable. As a basic formulation of the private search problem, consider a dataset $\Delta$  that is replicated across $N$ non-colluding servers. The dataset is comprised of $L$ i.i.d. records, $\Delta=(\Delta_1, \Delta_2, \cdots, \Delta_L)$. Each record $\Delta_l$ takes values uniformly in a set $\mathcal{U}=\{U_1, U_2,\cdots, U_K\}$. Private search allows a user to privately choose a $\theta\in[K]$ and then search for all records that satisfy $\Delta_l=U_\theta$, without revealing any information about $\theta$ to any individual server. Suppose $L\gg K\gg 1$, i.e., the alphabet size, $K$, is  large, but the number of records in the dataset is much larger. This is not an uncommon scenario. 
For example, consider datasets of DNA sequences. When searching for a DNA pattern of length $\ell$ (e.g., $\ell=10$), the alphabet size is $K=4^\ell$, while current DNA sequencing machines produce millions of records (called reads) per run.
Since the upload cost of private search does not depend on $L$ while the download cost scales linearly with $L$, the communication cost of private search for large $L$ is dominated by the download cost. The \emph{capacity} of private search is therefore defined as the maximum number of bits of desired information that can be retrieved per bit of download. Furthermore, since $K\gg 1$,  the \emph{asymptotic capacity} of private search, i.e., the capacity for large $K$ is of particular interest. Characterizing the asymptotic capacity of private search is our main goal in this work.

Private search (PS) as formulated above is closely related to private information retrieval (PIR). Recall that in its original form as introduced by Chor et al. in \cite{PIRfirstjournal}, the goal of PIR is to allow a user to retrieve an arbitrary desired message out of $\mu$ \emph{independent} messages that are replicated across $N$ distributed and non-colluding servers, without revealing any information about the identity of the desired message to any individual server. The capacity of PIR, $C_{\mbox{\scriptsize \it PIR}}(\mu,N)$,  is the fundamental limit on the number of bits of desired information that can be retrieved per bit of download. It was shown in \cite{Sun_Jafar_PIR} that $C_{\mbox{\scriptsize \it PIR}}(\mu,N)= (1+\frac{1}{N}+\cdots+\frac{1}{N^{\mu-1}})^{-1}$. The capacity of several variants of PIR has since been characterized \cite{Sun_Jafar_TPIR,Tajeddine_Rouayheb,Sun_Jafar_SPIR,Banawan_Ulukus,Wang_Skoglund,Tandon_CachePIR, Banawan_Ulukus_Multimessage, Sun_Jafar_PC, Mirmohseni_Maddah}. Particularly relevant to Private Search is the generalized form of PIR introduced in  \cite{Sun_Jafar_PC}, known as the Private Computation problem. As its main result, \cite{Sun_Jafar_PC} establishes the capacity of PIR when the messages have arbitrary linear dependencies. A supplementary result of \cite{Sun_Jafar_PC} shows that even if non-linear dependencies are allowed, the asymptotic capacity of Private Computation approaches $1-1/N$ provided that the message set  includes an unbounded number of independent messages. Private search is not covered by either result because in private search the dependencies among messages are non-linear and  no two messages are independent. To see this clearly, note that the search for all records that match $U_\theta$ is equivalent to retrieving the message $W_\theta, \theta\in[K],$ comprised of $L$ i.i.d. bits, $W_\theta = (W_\theta(1), W_\theta(2), \cdots, W_\theta(L))$, such that $W_\theta(l)=1$ if $\Delta_l=U_\theta$, and $W_\theta(l)=0$ otherwise. It is easily seen that any two messages, $W_i, W_j$, $i\neq j$, are identically distributed but not independent, e.g.,  $W_i(l)=1$ implies $ W_j(l)=0$. 
Therefore, we consider a broader generalization of PIR to include messages with arbitrary dependencies (DPIR in short). Private Search is  a special case of DPIR.


Our main contributions are as follows. We start with a general (non-asymptotic) converse for DPIR (Theorem \ref{theorem:alpha-bound}). Converse here denotes a lower bound on the download cost (equivalently, an upper bound on the capacity). Combined with a general achievability result for DPIR that was established in \cite{Sun_Jafar_PC}, this bound leads us to a sufficient condition (Theorem \ref{theorem:suf}) under which the asymptotic capacity of DPIR converges to $1-1/N$. The sufficient condition is  shown to hold for private search, thus establishing (Theorem \ref{theorem:PScap}) the asymptotic capacity of private search as $1-1/N$. 
As a natural generalization of private search, we consider $M$-approximate private search, where the user retrieves an arbitrary $L$ bit message $W_i$ such that the $l^{th}$ bit  of $W_i$ is equal to $1$ if $\Delta_l\in S_i=\{\theta_1, \theta_2, \cdots, \theta_M\}\subset[K]$, and $0$ otherwise. Note that there are $\mu=\binom{K}{M}$ possible messages, corresponding to $\binom{K}{M}$ possible choices of $S_i$. The sufficient condition of Theorem \ref{theorem:suf}  also holds for $M$-approximate private search, even when $M$ itself grows with $K$, so that the asymptotic capacity of approximate search (Theorem \ref{theorem:PScap}) is also equal to $1-1/N$. Finally, to illustrate the difficulty of finding general asymptotic capacity results for DPIR, we consider an example of approximate private search with restricted search patterns. For this example, we show (Proposition \ref{prop:special})  that either the new converse bound is not tight, or the asymptotic capacity is not $1-1/N$. The asymptotic capacity for this example remains open.

{\it Notation:} $[z_1:z_2]$ represents the set $\{z_1,z_1+1,\cdots, z_2\}$, for $z_1, z_2\in\mathbb{N}$, $z_1<z_2$,  $[Z]$ represents $[1:Z]$ for $Z\in\mathbb{N}$,  and for any set $S$, $W_{S}$ represents $\{W_i: i\in S\}$. $A \sim B$ means that random vectors $A$ and $B$ are identically distributed. A function $f(L)=o(L)$ means that $\lim_{L\rightarrow\infty}f(L)/L=0$. A function $f(L)=\Omega(L)$ means that $\lim_{L\rightarrow\infty}\left |f(L)\right |/L \ge c$, for some constant  $c>0$.

\section{Problem Statement}
\subsection{Dependent Private Information Retrieval (DPIR)}
Consider $\mu\in\mathbb{N}$ messages, $W_m, m\in[\mu]$, each comprised of $L$ i.i.d. symbols, $W_m=(W_m(1), W_m(2), \cdots, W_m(L))$, so that for each $l\in[L]$, the tuple $(W_1(l), W_2(l), \cdots, W_\mu(l))$ is an i.i.d. realization of the random $\mu$-tuple, $(w_1, w_2, \cdots, w_\mu)$. Thus, message realizations are i.i.d. across $l$, but for any particular $l$, the message symbols have dependencies defined by the joint distribution of $w_m, m\in[\mu]$. Also, $\forall m\in[\mu]$,
\begin{eqnarray}
H(W_m)&=&LH(w_m).
\end{eqnarray}
We say that the DPIR problem is \emph{balanced} if all messages carry the same amount of information,
\begin{eqnarray}
H(w_1)=H(w_2)=\cdots=H(w_\mu) \triangleq H(w).
\end{eqnarray}

There are $N$ servers and each server stores all the $\mu$ messages. A user privately generates $\theta \in [\mu]$ and wishes to retrieve $W_\theta$ while keeping $\theta$ a secret from each server. Depending on $\theta$, the user employs $N$ queries $Q_1^{[\theta]}, \cdots, Q_N^{[\theta]}$ and sends $Q_n^{[\theta]}$ to the $n^{th}$ server. The $n^{th}$ server returns a response string $A_{n}^{[\theta]}$ which is a function of $Q_{n}^{[\theta]}$ and ${W}_{[\mu]}$, i.e.,
\begin{eqnarray}
\forall \theta\in[\mu], \forall n\in[N],~H(A_n^{[\theta]} \mid Q_n^{[\theta]}, W_{[\mu]}) = 0.
\end{eqnarray}
From all the information that is now available to the user, he must be able to decode the desired message $W_\theta$, with probability of error $P_e$ which must approach zero as $L\rightarrow\infty$. This is called the ``correctness'' constraint. From Fano's inequality, we have 
\begin{eqnarray}
\mbox{[Correctness]} ~H\left({W}_{\theta} \mid A_{[N]}^{[\theta]},  Q_{[N]}^{[\theta]}\right) = o(L). \label{DPIR correct}
\end{eqnarray}

To protect the user's privacy, $\theta$ must be indistinguishable from $\theta'$, from the perspective of each server $\forall \theta, \theta'\in[\mu]$, i.e.,
\begin{eqnarray}
\label{DPIR privacy}
\mbox{[Privacy]} ~ (Q_n^{[\theta]}, A_n^{[\theta]}, {W}_{[\mu]}) \sim (Q_n^{[\theta']}, A_n^{[\theta']}, {W}_{[\mu]}).
\end{eqnarray}

The DPIR \emph{rate} characterizes how many bits of desired information are retrieved per downloaded bit, and is limited\footnote{If the DPIR problem is balanced, then the minimum over $m$ may be ignored.} by the worst case as, 
\begin{eqnarray}
R \triangleq \frac{\min_{m\in[\mu]}LH({w}_m)}{D}, \label{eq:rate}
\end{eqnarray}
where $D$ is the expected total number of bits downloaded by the user from all the servers. The supremum of achievable rates $R$ is the \emph{capacity} $C_{\mbox{\scriptsize \it DPIR}}(\mu, N)$.

\subsection{Private Search}
Consider a dataset ${\Delta}$ comprised of $L$ i.i.d. records: ${\Delta}=(\Delta_1, \Delta_2, \cdots, \Delta_L)$. Each record $\Delta_l$, $l\in[L]$, takes values uniformly from the alphabet set $\mathcal{U}=\{U_1, U_2, \cdots, U_K\}$. The dataset is replicated across $N$ non-colluding servers.
\begin{eqnarray}
P(\Delta_l=U_k)&=&\frac{1}{K},~~~~~~\forall l\in[L],k\in[K],\\
H({\Delta})&=&LH(\Delta_l) = L\log_2(K)\mbox{ bits}.
\end{eqnarray}
A user privately chooses a  set, $S=\{U_{\theta_1}, U_{\theta_2}, \cdots, U_{\theta_M}\}$,  $S\subset\mathcal{U}$, $M<K$, and searches for all records in $\Delta$ that match any of the elements of $S$. We refer to the $M=1$ setting as \emph{exact} private search, and to the $M>1$ setting as \emph{approximate} private search (or $M$-approximate private search), because the output of the search reveals the exact value of a matching record if $M=1$, but not if $M>1$. In general, for approximate search we allow $M$ to grow with $K$ (either $o(K)$ or $\Omega(K)$) in the asymptotic regime $K\rightarrow\infty$.

There are a total of $\mu=\binom{K}{M}$ possible choices of the search set $S$. Let us arbitrarily label them $S_m, m\in[\mu]$. Correspondingly, there are a total of $\mu$ messages for $M$-approximate private search. Label these messages $W_m$, so that $\forall m\in[\mu]$,
\begin{eqnarray}
W_m=(W_m(1), W_{m}(2), \cdots, W_{m}(L)),
\end{eqnarray}
and
\begin{eqnarray} \label{eq:exact}
W_{m}(l)=\left\{
\begin{array}{ll}
1,& \Delta_l\in S_m,\\
0& \mbox{otherwise}.
\end{array}
\right.
&&\forall l\in[L].
\end{eqnarray}
Note that the $L$ bits of each message are i.i.d.
\begin{eqnarray}
H(w)=H(W_{m}(l))=H_2\left(M/K\right), \forall l\in[L], \forall m\in[\mu],\nonumber
\end{eqnarray}
where the  binary entropy function is defined as follows.
\begin{eqnarray}
H_2(p)=-p\log_2(p)-(1-p)\log_2(1-p),
\end{eqnarray}
$H_2(0)=H_2(1)=0$. 

%
%
%
%

The queries and answers, privacy and correctness constraints, rate and capacity definitions for private search are inherited from DPIR. The \emph{capacity of private search} is denoted $C_{\mbox{\scriptsize \it PS}}(K,M,N)$, and the asymptotic capacity of private search is denoted $\lim_{K\rightarrow\infty}C_{\mbox{\scriptsize \it PS}}(K,M,N)$.

\section{Results}
We present the main results in this section. All proofs appear in Section \ref{sec:proofs}.
\subsection{A General Converse for DPIR}
The download cost (expected number of bits of download) for DPIR is bounded as follows.
\begin{theorem}\label{theorem:alpha-bound}
For DPIR, denote by $W_1,W_2,\dots,W_\mu$ an arbitrary permutation of the $\mu$ messages. Then
\begin{eqnarray}
D\geq H(W_1)+\frac{H(W_2|W_1)}{N}+\cdots+\frac{H(W_\mu|W_{[\mu-1]})}{N^{\mu-1}}.
\label{eq:alpha-bound}
\end{eqnarray}
\end{theorem}

Note that the bound depends on the chosen permutation of message indices, so finding the best bound from Theorem \ref{theorem:alpha-bound} requires a further optimization of the permutation. Substituting \eqref{eq:alpha-bound} into \eqref{eq:rate}, we obtain an equivalent bound on capacity. If the messages are independent, we recover the converse bound of \cite{Sun_Jafar_PIR}. However, Theorem \ref{theorem:alpha-bound} is more broadly useful since it allows arbitrary dependencies. Also note that Theorem \ref{theorem:alpha-bound} is not limited to balanced DPIR. 

\subsection{General Achievable Rate for DPIR \cite{Sun_Jafar_PC}}
The following achievable rate for DPIR is shown in \cite{Sun_Jafar_PC}.
\begin{theorem}\label{lemma:ach rate}(\cite{Sun_Jafar_PC}, Section $7$)
\begin{eqnarray}
C_{\mbox{\tiny DPIR}}(\mu,N)  \geq \left(1 - \frac{1}{N}\right)\frac{H_{\min}}{H_{\max}} \label{eq:generalrate}
\end{eqnarray}
where $H_{\min}=\min_{m\in[\mu]} H(w_m)$ and  $H_{\max}=\max_{m\in[\mu]} H(w_m)$.
\end{theorem}
For balanced DPIR, this gives us  $1-1/N$ as a lower bound on capacity.

\subsection{Asymptotic Optimality of Rate $1-1/N$ for Balanced DPIR}
For balanced DPIR, as the number of messages $\mu\rightarrow\infty$,  the asymptotic behavior of \eqref{eq:alpha-bound} gives us the following sufficient condition. Here we define $W_k=0$ if $k>\mu$.
\begin{theorem} \label{theorem:suf}
For balanced DPIR, if there exists an increasing sequence $k_i\in\mathbb{N}, \forall i\in\mathbb{N}$, such that $\forall l\in\mathbb{N}$,
\begin{eqnarray}
\lim_{\mu\rightarrow\infty}\frac{I\left(W_{k_{l+1}}; W_{k_{[1:l]}}\right)}{LH(w)}=0,
 \label{eq:guessone}
\end{eqnarray}
then the asymptotic capacity is 
\begin{eqnarray}
\lim_{\mu \rightarrow \infty}C_{\mbox{\tiny DPIR}}(\mu,N) = 1-\frac{1}{N}.
\end{eqnarray}
\end{theorem}
Note since $H(w)$ may depend on $\mu$, the sufficient condition is in general not equivalent to $\lim_{\mu\rightarrow\infty}I\left(W_{k_{l+1}}; W_{k_{[1:l]}}\right)=0$.

\subsection{Asymptotic Capacity of Private Search}
\begin{figure}[!htbp]
   \centering
   \includegraphics[width=0.3\textwidth]{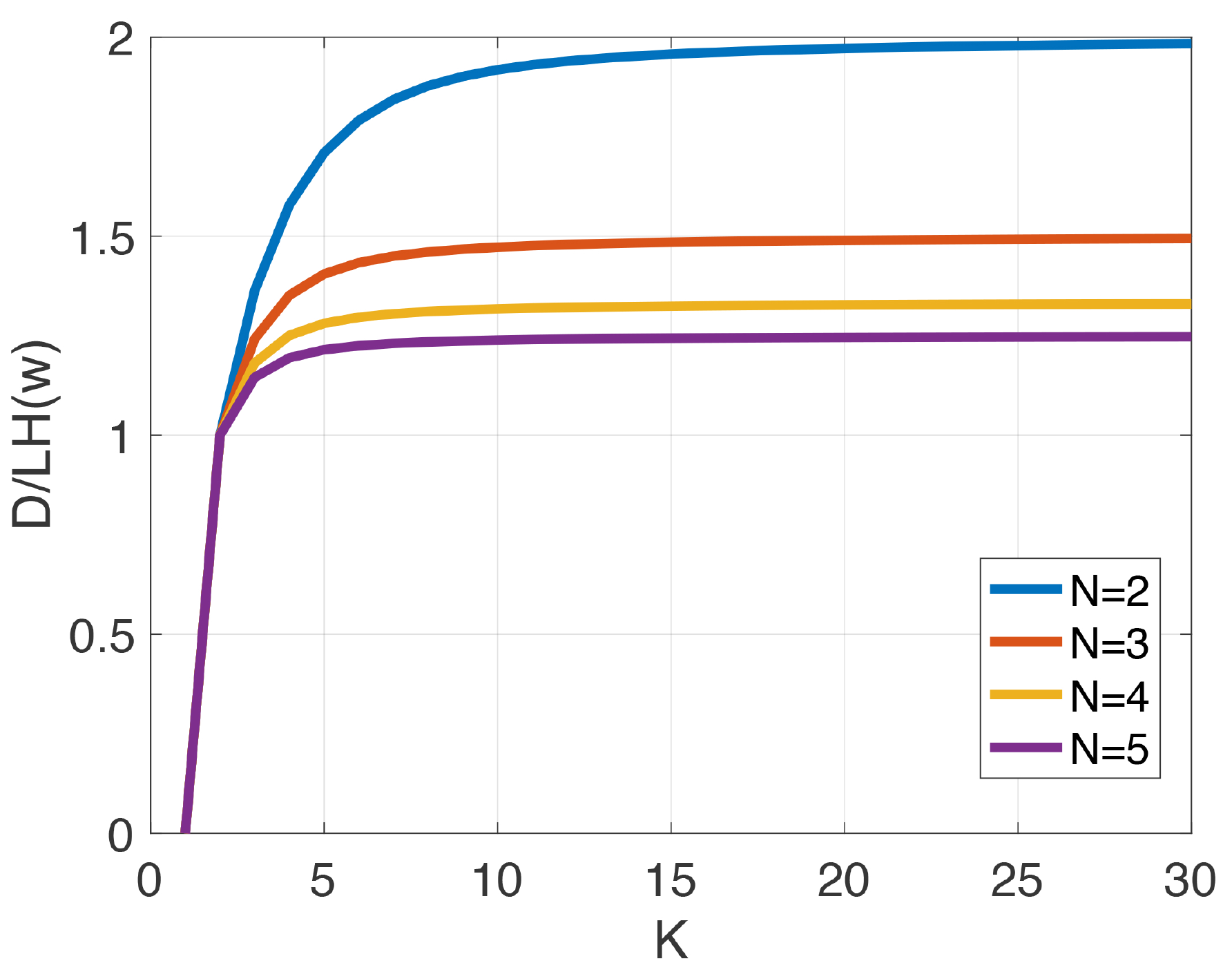}
   \caption{Normalized download lower bound of exact search based on Theorem 1
versus alphabet size $K$. The asymptotic value $(1-1/N)^{-1}$ is the upper
bound.}
   \label{fig}
\end{figure}

\begin{theorem} \label{theorem:PScap}
The asymptotic capacity of private search is
\begin{eqnarray}
\lim_{K\rightarrow\infty} C_{\mbox{\scriptsize \it PS}}(K, M, N)&=& 1-\frac{1}{N},
\end{eqnarray}
for both exact private search $(M=1)$ and approximate private search $(M>1)$.
\end{theorem}
Theorem \ref{theorem:PScap} is proved by showing that the sufficient condition \eqref{eq:guessone} is satisfied for private search. Notably, for exact private search, as $K\rightarrow\infty$, both $I(W_{k_{l+1}}; W_{k_{[1:l]}})$ and $H(w)$ approach zero. The key to the asymptotic capacity result is that $I(W_{k_{l+1}}; W_{k_{[1:l]}})$ approaches zero much faster than $H(w)$. Furthermore, as shown in Fig. \ref{fig}, convergence of capacity to its asymptotic value is quite fast, and the larger the value of $N$, the faster the convergence. For example, with $N=5$, the bound (\ref{eq:alpha-bound}) for $K=10$ messages is already within $1\%$ of the asymptotic value.

\subsection{Difficulty of Private Search over Restricted Search Patterns}
Finding the capacity of DPIR with arbitrary dependency structures is in general a difficult problem. The difficulty remains even when the problem is limited to asymptotic capacity.  To highlight this aspect, we present an example of approximate private search over restricted search patterns where the asymptotic capacity remains an open problem.
\begin{proposition} \label{prop:special}
Consider $M$-approximate private search, with $M=\lfloor\frac{K}{2}\rfloor$, where the only search sets allowed are
\begin{eqnarray}
S_k=\{U_{<k+1>}, U_{<k+2>}, \cdots, U_{<k+M>}\}, ~~\forall k\in[K],
\end{eqnarray}
and   $<m>=m \mod K + 1$. As $K\rightarrow\infty$, either the bound \eqref{eq:alpha-bound} is not tight, or $\lim_{K\rightarrow\infty} C_{\mbox{\scriptsize \it PS}}(K, M, N)\neq 1-\frac{1}{N}$.
\end{proposition}
Here privacy is required only within the $\mu=K$ choices.

\section{Proofs}\label{sec:proofs}
\subsection{Proof of Theorem \ref{theorem:alpha-bound}}
For the DPIR problem, the total download is bounded as,  
\begin{eqnarray}
D &\geq& H(A_{[N]}^{[1]}\mid Q^{[1]}_{[N]}) \overset{(\ref{DPIR correct})}{=} H(A_{[N]}^{[1]}, {W}_{1}\mid Q^{[1]}_{[N]})\\
&=& H({W}_{1}\mid Q^{[1]}_{[N]})+H(A_{[N]}^{[1]} \mid Q^{[1]}_{[N]}, {W}_{1})\\
&\geq& H({W}_{1})+H(A_1^{[1]} \mid Q^{[1]}_{[N]}, {W}_{1})\\
&=& H({W}_{1})+H(A_1^{[1]} \mid {Q}_1^{[1]}, {W}_{1})\\
&\overset{(\ref{DPIR privacy})}{=}& H({W}_{1})+H(A_1^{[2]}\mid {Q}_1^{[2]}, {W}_{1})\\
&=& H({W}_{1})+H(A_1^{[2]}\mid Q^{[2]}_{[N]}, {W}_{1}).
\end{eqnarray}
Similarly, for all $n\in[2:N]$ we have,
\begin{eqnarray}
D &\geq& H({W}_{1}) + H(A_n^{[2]}\mid Q^{[2]}_{[N]}, {W}_{1}).
\end{eqnarray}
Adding all of these $N$ inequalities we obtain,
\begin{eqnarray}
D &\geq& H({W}_{1}) + \frac{H(A_{[N]}^{[2]}\mid Q^{[2]}_{[N]},{W}_{1})}{N} .
\end{eqnarray}

\noindent Proceeding along the lines of the recursive proof of \cite{Sun_Jafar_PIR},
\small
\begin{eqnarray}
D &\geq& H({W}_{1}) + \frac{H(A_{[N]}^{[2]}\mid Q^{[2]}_{[N]}, {W}_{1})}{N} \\
&\geq& \cdots \nonumber\\
&\geq& H({W}_{1}) + \frac{H({W}_{2} \mid {W}_{1})}{N} + \cdots + \frac{H({W}_{{\mu}} \mid {W}_{{[\mu-1]}})}{N^{\mu-1}}.
\end{eqnarray}
\normalsize 

\subsection{Proof for Theorem \ref{theorem:suf}}
Define $m$ such that $k_m \leq \mu < k_{m+1}$. Note that $m$ is a function of $\mu$ and that as $\mu \rightarrow \infty$,  $m\rightarrow\infty$.  Based on Theorem \ref{theorem:alpha-bound}, the normalized download is bounded as  
\small
\begin{eqnarray}
\frac{D}{LH(w)} &\geq& \frac{H({W}_{k_1})}{LH(w)} + \cdots +  \frac{H({W}_{k_{m}} \mid {W}_{k_{[1:m-1]}})}{N^{m-1}LH(w)} \nonumber\\
&\geq& (1+\frac{1}{N}+\frac{1}{N^2}+\cdots+\frac{1}{N^{m-1}}) \\
&&-\frac{I({W}_{k_2}; {W}_{k_{1}})}{NLH(w)}-\cdots-\frac{I({W}_{k_m}; {W}_{k_{[1:m-1]}})}{N^{m-1}LH(w)}.\nonumber
\end{eqnarray}
\normalsize
Applying limit $\mu\rightarrow\infty$, the reciprocal of rate is bounded as
\small
\begin{eqnarray}
\lim_{\mu\rightarrow\infty}\frac{D}{LH(w)}&\geq&\left(1-\frac{1}{N}\right)^{-1}-\lim_{\mu\rightarrow\infty}\sum_{l=1}^{m-1}\frac{I({W}_{k_{l+1}};{W}_{k_{[1:l]}})}{LH(w) N^l}. \nonumber
\end{eqnarray}
\normalsize
Now, we need to show that 
\begin{eqnarray}
\lim_{\mu\rightarrow\infty}\sum_{l=1}^{m-1}\frac{I({W}_{k_{l+1}};{W}_{k_{[1:l]}})}{LH(w) N^l}&=&0. \label{eq:claim}
\end{eqnarray}
Equivalently, for every $\epsilon>0$ we will show that 
\begin{eqnarray}
\lim_{\mu\rightarrow\infty}\sum_{l=1}^{m-1}\frac{I({W}_{k_{l+1}};{W}_{k_{[1:l]}})}{LH(w) N^l}&\leq&\epsilon.
\end{eqnarray}
Choose a finite $l^*$ such that
\begin{eqnarray}
\frac{1}{N^{l^*}}\left(1-\frac{1}{N}\right)^{-1}\leq {\epsilon}.
\end{eqnarray}
Note that $l^*$ depends only on $N$ and $\epsilon$. More importantly, it is not a function of $\mu$. Now partition the sum as follows
\begin{eqnarray}
\lim_{\mu\rightarrow\infty}\sum_{l=1}^{m-1}\frac{I({W}_{k_{l+1}};{W}_{k_{[1:l]}})}{LH(w) N^l} 
=\lim_{\mu \rightarrow\infty}\sum_{l=1}^{l^*-1}\frac{I({W}_{k_{l+1}};{W}_{k_{[1:l]}})}{LH(w) N^l}\nonumber\\
+
\lim_{\mu \rightarrow\infty}\sum_{l=l^*}^{m-1}\frac{I({W}_{k_{l+1}};{W}_{k_{[1:l]}})}{LH(w) N^l}.~~~~~
\label{eq:two}
\end{eqnarray}
The first term on the RHS of \eqref{eq:two} is zero because it is a sum of finitely many terms ($l^*$ is finite), each of which is zero because \eqref{eq:guessone} holds by assumption. For the second term in (\ref{eq:two}),
\begin{eqnarray}
\lim_{\mu \rightarrow\infty}\sum_{l=l^*}^{m-1}\frac{I({W}_{k_{l+1}};{W}_{k_{[1:l]}})}{LH(w) N^l}
\leq \lim_{\mu \rightarrow\infty}\sum_{l=l^*}^{m-1}\frac{1}{N^l}\\
\leq \frac{1}{N^{l^*}}\lim_{\mu \rightarrow\infty}\sum_{l=0}^{\mu }\frac{1}{N^l}
\leq \frac{1}{N^{l^*}}\left(1-\frac{1}{N}\right)^{-1}
\leq {\epsilon}. \label{eps1}
\end{eqnarray}
Thus, the reciprocal of rate is bounded as $1/R \geq (1-1/N)^{-1}$, i.e., the rate is bounded as $R\leq 1-1/N$. By Theorem \ref{lemma:ach rate} this rate is achievable. Hence proved.

\subsection{Proof of Theorem \ref{theorem:PScap}}
Let us start with exact private search $(M = 1)$.
\begin{eqnarray}
\lim_{K\rightarrow\infty}H(w)=\lim_{K\rightarrow\infty}H_2\left(\frac{1}{K}\right)=0.
\end{eqnarray}
Choosing the sequence $k_i=i$ and substituting $\mu=K$ into the LHS of (\ref{eq:guessone}), we have
\begin{eqnarray}
&& \lim_{K\rightarrow\infty}\frac{I(W_{l+1}; W_{1}, W_{2}, \cdots, W_{l})}{LH(w)}\nonumber\\
 &=&\lim_{K\rightarrow\infty}\frac{H_2\left(\frac{1}{K}\right)-\left(1-\frac{l}{K}\right)H_2\left(\frac{1}{K-l}\right)}{H_2\left(\frac{1}{K}\right)} \\
 &=&1 - \lim_{K\rightarrow\infty}\frac{\left(1-\frac{l}{K}\right)H_2\left(\frac{1}{K-l}\right)}{H_2\left(\frac{1}{K}\right)}=0.
\end{eqnarray}
Therefore, (\ref{eq:guessone}) is satisfied, and based on Theorem \ref{theorem:suf}, the asymptotic capacity of exact private search is $1 - 1/N$.

For approximate search $(M>1)$, define $\gamma \triangleq M/K<1$. By symmetry of the truth function, we only consider $\gamma \leq 1/2$.

When $M = o(K)$, consider messages with disjoint patterns. For example, the alphabet set is $\{1,2,\cdots,K\}$, $M=2$. Consider messages corresponding to $\{1,2\}, \{3,4\}, \{5,6\},\cdots$. Since these patterns are disjoint, they can be viewed as $M=1$ and alphabet size of $K/2$. As $K \rightarrow \infty$, the number of messages $K/M \rightarrow \infty$. Then, the asymptotic capacity is $1-1/N$ in this case.

For $M = \Omega(K)$, let us find a sequence of dependent messages such that (\ref{eq:guessone}) is satisfied.  Choose  $W_1$ corresponding to $S_1=\{U_1,U_2,\cdots,U_M\}$. It separates the alphabet set into $2$ parts: $S_1$ of size $\gamma K$,  and $\mathcal{U}\backslash S_1$ of size $(1-\gamma)K$. Note that $\gamma K=M$ is an integer. Choose the second message $W_2$ so that it is comprised of $\floor{\gamma M}$ elements of $S_1$ and $M-\floor{\gamma M}$ elements of $\mathcal{U}\backslash S_1$. Repeating this step we  get a series of dependent messages. Let us represent $U_1, U_2, \cdots, U_K$ on an alphabet line $\mathcal{U}$ as follows.\\

\begin{centering}
\begin{tikzpicture}
\draw [black] (0,2) -- (6,2);
\draw [red, ultra thick] (0, 1.4) -- (2,1.4);
\draw [black] (2,1.4) -- (6,1.4);
\draw [blue, ultra thick] (0, 0.8) -- (2/3,0.8);
\draw [black] (2/3,0.8) -- (2,0.8);
\draw [blue, ultra thick] (2, 0.8) -- (10/3,0.8);
\draw [black] (10/3,0.8) -- (6,0.8);
\draw [green, ultra thick] (0, 0.2) -- (2/9,0.2);
\draw [black] (2/9,0.2) -- (6/9,0.2);
\draw [green, ultra thick] (6/9, 0.2) -- (10/9,0.2);
\draw [black] (10/9,0.2) -- (18/9,0.2);
\draw [green, ultra thick] (18/9, 0.2) -- (22/9,0.2);
\draw [black] (22/9, 0.2) -- (10/3,0.2);
\draw [green, ultra thick] (30/9, 0.2) -- (38/9,0.2);
\draw [black] (38/9,0.2) -- (6,0.2);
\node at (-0.5,2) {\scriptsize $\mathcal{U}$};
\node at (-0.5,1.4) {\scriptsize $W_1$};
\node at (-0.5,0.8) {\scriptsize $W_2$};
\node at (-0.5,0.2) {\scriptsize $W_3$};
\node at (1,1.7) {\scriptsize $M$};
\node at (1/3,1.1) {\scriptsize $\floor{\gamma M}$};
\node at (8/3,1.1) {\scriptsize $M-\floor{\gamma M}$};
\node at (1/3,0.5) {\scriptsize $\floor{\gamma \floor{\gamma M}}$};
\node at (8/9,-.1) {\scriptsize $\floor{\gamma M-\gamma\floor{\gamma M}}$};
\node at (20/9,0.5) {\scriptsize $\floor{\gamma M-\gamma\floor{\gamma M}}$};
\node at (48/9,-.1) {\scriptsize $M-2\floor{\gamma M-\gamma\floor{\gamma M}}-\floor{\gamma\floor{\gamma M}}$};
\node at (3,-.1) {\large $\cdot$};
\node at (3,-.3) {\large $\cdot$};
\node at (3,-.5) {\large $\cdot$};

\foreach \dis [count=\xi] in {0, 0.5, ..., 6}
{
  \node[font=\large] at (\dis, 2) {\tiny $\bullet$};
}
\node at (0,2.3){\scriptsize $U_1$};
\node at (0.5,2.3){\scriptsize $U_2$};
\node at (1.0,2.3){\scriptsize $U_3$};
\node at (3.5,2.3){\scriptsize $\cdots$};
\node at (6,2.3){\scriptsize $U_K$};
\end{tikzpicture}\\
\end{centering}

Note that
\begin{eqnarray}
 H(W_l) = LH_2(\gamma),~\forall l .
\end{eqnarray}
\footnotesize
\begin{eqnarray}
&H(W_2|W_1) = LH_2\left(\frac{\floor{\gamma M}}{M}\right) \frac{M}{K} + LH_2\left(\frac{M-\floor{\gamma M}}{K-M}\right) \frac{K-M}{K}  \\
&\geq LH_2\left(\frac{\gamma M-1}{M}\right) \frac{M}{K} + LH_2\left(\frac{\gamma(K-M)-1}{K-M}\right) \frac{K-M}{K} \label{eq:non}\\
&= LH_2\left(\frac{\gamma^2K-1}{\gamma K}\right) \frac{M}{K} + LH_2\left(\frac{\gamma(1-\gamma)K-1}{(1-\gamma)K}\right) \frac{K-M}{K} \\
&\Rightarrow\lim_{K\rightarrow \infty}H(W_2|W_1) \geq LH_2(\gamma)= H(W_1).\end{eqnarray}
\normalsize
One can show that even when $\frac{M-\floor{\gamma M}}{K-M}$ and $\frac{\gamma(K-M)-1}{K-M}$ are in non-monotonic range, (\ref{eq:non}) is still true. 
Since $M=\Omega(K)$, there exists a constant $0<c<1$ such that $\gamma=M/K \ge c$ for sufficiently large $K$. For a given $K$, consider the search of only messages $\{W_l: l \leq \log_{1/c}\sqrt{K} \}$. Note that the number of messages goes to infinity as $K \to \infty$.
Next we prove
\small
\begin{eqnarray}
\lim_{K\rightarrow \infty}\frac{H(W_l | W_1, \cdots, W_{l-1})}{LH_2(\gamma)} = 1,~\forall l \leq \log_{1/c}\sqrt{K}.
\end{eqnarray}
\normalsize
Based on the construction above, 
there are $2^{l-1}$ terms in $H(W_l | W_1,\cdots, W_{l-1})$. To bound the $i^{th}$ term, first let us use a binary number to represent $i-1$.  Let the number of ``1''s in the binary number be $m_i$. For example, if $l=4$ and $i=2$, then $i-1=(001)_2$, and $m_i=1$. Using a similar argument as for $l=2$, we can partition the alphabet into $2^l$ parts at step $l$. The size of the $i^{th}$ part is between $\gamma^{l-m_i}(1-\gamma)^{m_i}K-l+1$ and $\gamma^{l-m_i}(1-\gamma)^{m_i}K+l-1$. Then the $i^{th}$ term of $H(W_l | W_1,\cdots, W_{l-1})$ is greater than or equal to 
\begin{eqnarray}
&& LH_2\left(\frac{\gamma^{l-m_i+1}(1-\gamma)^{m_i}K-l+1}{\gamma^{l-m_i}(1-\gamma)^{m_i}K+l-1}\right) \times P(\mbox{$i^{th}$ term})\nonumber\\
&&= LH_2\left(\frac{\gamma-\frac{l-1}{\gamma^{l-m_i}(1-\gamma)^{m_i}K}}{1+\frac{l-1}{\gamma^{l-m_i}(1-\gamma)^{m_i}K}}\right) \times P(\mbox{$i^{th}$ term})
\end{eqnarray}
When $K\rightarrow \infty$, $\forall i \in [l]$, $l \leq \log_{1/c}\sqrt{K}$,
\begin{eqnarray}
\lim_{K\rightarrow \infty}\frac{l-1}{\gamma^{l-m_i}(1-\gamma)^{m_i}K} \leq \lim_{K\rightarrow \infty}\frac{l-1}{\gamma^lK} = 0.
\end{eqnarray}
Therefore,
\begin{eqnarray}
\lim_{K\rightarrow \infty}LH_2\left(\frac{\gamma-\frac{l-1}{\gamma^{l-m_i}(1-\gamma)^{m_i}K}}{1+\frac{l-1}{\gamma^{l-m_i}(1-\gamma)^{m_i}K}}\right) =\lim_{K\rightarrow \infty} LH_2(\gamma).
\end{eqnarray}
Summing up all the terms, we obtain 
\footnotesize
\begin{eqnarray}
\lim_{K\rightarrow \infty} H(W_l | W_1,\cdots, W_{l-1}) \geq \lim_{K\rightarrow \infty} LH_2(\gamma)
= \lim_{K\rightarrow \infty} H(W_l).
\end{eqnarray}
\normalsize
Invoking Theorem \ref{theorem:suf} at this point, we conclude that the asymptotic capacity is $1-1/N$. Hence proved.

\subsection{Proof of Proposition \ref{prop:special}} \label{sec:rough}
Consider the even values of $K$ as it approaches infinity, so that we have $H(W_k(l))=H(1/2)=1$ bit, i.e., each message bit is marginally uniform. Suppose there exists a sequence $k_1, k_2, \cdots$ for which according to Theorem \ref{theorem:suf}, $\lim_{K\rightarrow\infty}D/LH_2(1/2)=(1-1/N)$. Then the following must hold.
\begin{eqnarray}
\lim_{K\rightarrow\infty}H\left(W_{{k_2}}(l)\mid W_{{k_1}}(l)\right)&=&1\label{eq:deal1}\\
\lim_{K\rightarrow\infty}H\left(W_{{k_3}}(l)\mid W_{{k_1}}(l), W_{{k_2}}(l)\right)&=&1\label{eq:deal2}
\end{eqnarray}
Represent $U_1, U_2, \cdots, U_K$ on an alphabet circle $\mathcal{U}$ as follows.

\begin{centering}
\begin{tikzpicture}[scale=0.75]
\draw [black] (0,0) circle [radius=2cm];
\foreach \angle [count=\xi] in {75,60,...,-270}
{
  \node[font=\large] at (\angle:2cm) {\tiny $\bullet$};
}
\node at (-235:1.7cm){\footnotesize $\cdot$};
\node at (-240:1.7cm){\footnotesize $\cdot$};
\node at (-230:1.7cm){\footnotesize $\cdot$};
\node at (-255:1.7cm){\tiny $U_K$};
\node at (90:1.7cm){\tiny $U_1$};
\node at (75:1.7cm){\tiny $U_2$};
\node at (60:1.7cm){\tiny $U_3$};
\node at (45:1.7cm){\footnotesize $\cdot$};
\node at (40:1.7cm){\footnotesize $\cdot$};
\node at (35:1.7cm){\footnotesize $\cdot$};
\node at (0,0) {\footnotesize $\mathcal{U}$};
\node at (-90:1.2cm) {\footnotesize A};
\node at (0:1.2cm) {\footnotesize B};
\node at (90:1.2cm) {\footnotesize C};
\node at (180:1.2cm) {\footnotesize D};
\draw [ultra thick, red] ([shift={(0,0)}]50:2.2) arc[radius=2.2, start angle=50, end angle= 230];
\node [red] at (160:3.0cm){\footnotesize $S_{k1}$};
\draw [ultra thick, blue] ([shift={(0,0)}]-40:2.4) arc[radius=2.4, start angle=-40, end angle= 140];
\node [blue] at (50:3.0cm){\footnotesize $S_{k2}$};
\end{tikzpicture}\\
\end{centering}

Since $S_{k_1}$ is a contiguous set of $K/2$ points on the circle, without loss of generality it may be represented by the red semi-circle. $W_{S_{k_1}}(l)$ and $W_{S_{k_2}}(l)$ are binary random variables. Since $\lim_{K\rightarrow\infty}H\left(W_{S_{k_2}}(l)\mid W_{S_{k_1}}(l)\right)=1$, within $S_{k_1}$ half of the points must be in $S_{k_2}$ and half of the points must be outside $S_{k_2}$. The same is true for the points outside $S_{k_1}$. Therefore, without loss of generality, $S_{k_2}$ is represented by the blue semi-circle on the alphabet circle. Note that this divides the alphabet circle into $4$ parts, labeled as $A, B, C, D$, corresponding to $(W_{S_{k_1}}(l), W_{S_{k_2}}(l))=(0,0), (0,1), (1,1), (1,0)$ respectively. 

Since $\lim_{K\rightarrow\infty}H\left(W_{S_{k_3}}(l)\mid W_{S_{k_1}}(l), W_{S_{k_2}}(l)\right)=1$, half of $A$ must be in $S_{k_3}$ and half of $A$ must be outside $S_{k_3}$. Similarly,  half of $B$, $C$, $D$ must be in $S_{k_3}$ and half of $B, C, D$ must be outside $S_{k_3}$. But $S_{k_3}$ is a contiguous semicircle, a continuous semi-circle cannot overlap with half of each of $A, B, C, D$. Therefore we have a contradiction.  The contradiction means that for this problem, either the asymptotic capacity of private search is not equal to $1-1/N$ or Theorem \ref{theorem:suf} is not tight. 

\section{Conclusion}
We introduced the private search problem, which requires PIR with dependent messages (DPIR). We derived a general converse bound for DPIR, studied its asymptotic behavior, and combined it with a known general achievability result in order to characterize the asymptotic capacity of private search. We also showed through an example that even asymptotic capacity characterizations for private search are difficult for additionally constrained message structures.

%

\IEEEtriggeratref{15}


\bibliographystyle{IEEEtran}
\bibliography{Thesis}

%
%
%
%
%
%

\end{document}